\documentclass[12]{article}
\usepackage{graphics,graphicx,amsbsy,amssymb,color}

\newcommand{\rr}{{\boldmath \mbox{$r$}}}

\newcommand{\vv}{{\boldmath \mbox{$v$}}}

\newlength{\defbaselineskip}
\newcommand{\setlinespacing}[1]%
           {\setlength{\baselineskip}{#1 \defbaselineskip}}

\title{\textbf{
Thermodynamics and Rate Thermodynamics }}
\author{Miroslav Grmela \\
\'{E}cole Polytechnique de Montr\'{e}al,
  C.P.6079 suc. Centre-ville,\\
 Montr\'{e}al, H3C 3A7,  Qu\'{e}bec, Canada}

 \date{}

\begin{document}

\maketitle

\tableofcontents
\setlength{\parskip}{4mm}

\begin{abstract}

Approach of mesoscopic state variables to time independent  equilibrium sates (zero law of thermodynamics) gives birth to the classical equilibrium thermodynamics.
Approach of  fluxes and forces to fixed points (equilibrium fluxes and forces)  that drive  reduced mesoscopic dynamics gives birth to the rate thermodynamics that is applicable to driven systems.
We formulate the rate thermodynamics and dynamics,  investigate its relation to the classical thermodynamics, to extensions involving more details,
to the hierarchy reformulations of dynamical theories, and   to the Onsager variational principle. We also compare thermodynamic and dynamic critical behavior  observed in  closed and open systems.  Dynamics and thermodynamics of the van der Waals gas provides an illustration.

\end{abstract}

\section{Introduction}\label{Intr}

The essential difference between externally unforced and externally driven systems is in the way they  approach reduced descriptions. In the former it is a gradual reduction that ends on the level of equilibrium thermodynamics on  which no time evolution takes place. This experimental observation is called  \textit{zero law of thermodynamics}.  In the latter the gradual reduction stops on a mesoscopic level on which the system still continues to evolve. The difference disappears if we  regard the reduction process in externally driven systems in the space of vector fields. The vector fields describing  externally driven systems approach  a time independent vector field that drives the reduced mesoscopic time evolution. No   further  reduction  to a simpler description is possible. With this observation (we call it a \textit{zero law of rate thermodynamics}) we can formulate rate thermodynamics (applicable to externally driven systems)   in the same way as the classical classical thermodynamics is formulated for externally unforced systems. We use hereafter the adjective "rate" to denote quantities appearing in investigations of dynamics and thermodynamics in the space of vector fields.

The approach of externally unforced systems to equilibrium states is driven by gradient of a potential, called entropy. The entropy, evaluated at the equilibrium state,  becomes the  entropy (the equilibrium entropy) that, if seen as a function of the total energy and the total number of moles (both constants of motion), provides the fundamental relation of equilibrium thermodynamics.

In the same way, but now in the space of vector fields, we arrive at the rate   thermodynamics. Its  main new feature (beside the fact that the state variables are rates of the state variables in the classical thermodynamics)  is that it is a thermodynamical theory applicable also
to externally driven systems.  The approach to the vector field generating the reduced mesoscopic dynamics is driven by gradient of the rate entropy. This entropy, if evaluated in the asymptotically reached vector field, becomes the entropy playing the same role in the rate thermodynamics as the entropy plays in the classical thermodynamics.

In order to outline the investigation presented in this paper and also in order to place it inside nonequilibrium thermodynamics and statistical mechanics we introduce three formal dynamical systems. In the course of our  investigation  they gradually  acquire  specific structures. Since we want our results to retain its meaning and applicability  on different scales (on different levels of description)  we try to minimize the structure. In other words, we try to give our analysis a multiscale character. An example of all three dynamical systems in which all symbols have a specific meaning is in Section \ref{vdW}.

The state variable in the first dynamical system is denoted by the symbol $x$, its time evolution is governed by
\begin{equation}\label{hyd10}
\frac{\partial x}{\partial t}=\mathbb{H}(J,x)|_{J=\mathfrak{J}(x)}
\end{equation}
The symbol $\mathbb{H}$ denotes at this point an unspecified operator,
$J$ represents another state variable that is however  specified as a function of $x$ by $J=\mathfrak{J}(x)$. For example, (\ref{hyd10}) can be a formal representation of the governing equations of hydrodynamics, $x$ are hydrodynamic fields and $J$ are either fluxes or variables characterizing the internal structure of the fluid under investigation. The specification $J=\mathfrak{J}(x)$ is called a constitutive relation.

The state variables in the second dynamical system are $(x,J)$, the time evolution equations are
\begin{equation}\label{zeta}
\frac{\partial}{\partial t}\left(\begin{array}{cc}x\\J\end{array}\right)=\mathbb{E}\mathbb{H}\left(\begin{array}{cc}x\\J\end{array}\right)
\end{equation}
The symbol $\mathbb{E}\mathbb{H}$ denotes at this point again an unspecified operator. The mesoscopic dynamical theory represented by (\ref{zeta}) includes more details than the theory represented by (\ref{hyd10}) due to the enlargement of the state space. The  extra state variable $J$ characterizes details that are not seen in the theory represented by (\ref{hyd10}). The simplest example of (\ref{zeta}) are governing equations of classical mechanics
$\frac{d}{d t}\left(\begin{array}{cc}r\\J\end{array}\right)=\left(\begin{array}{cc}J^*\\-r^*-aJ^*\end{array}\right)$ where $r$ is the position vector, its conjugate $r^*$ multiplied by $-1$ is the force,   $J$ is the momenum, its conjugate $J^*$ is the velocity. The term $-aJ^*$, where $a>0$ is a paranmetr, represnts the friction. Inspired by the terminology used in this example we shall call $J$ appearing in (\ref{zeta}) a momentum and $J^*$ a flux.
Another example of (\ref{zeta}) is the Grad hierarchy reformulation of the Boltzmann kinetic equation governing the time evolution of the one particle distribution function $f(\rr,\vv)$, $\rr$ is the position vector and $\vv$ the momentum of one particle. The state variable $x$ represents  the first five moments of $f(\rr,\vv)$ in $\vv$ and the momentum $J$ is the infinite number of remaining moments.

The third mesoscopic dynamical theory is  represented by
\begin{eqnarray}\label{clhi}
\frac{\partial x}{\partial t}&=&\mathbb{H}(J,x)|_{J=\mathfrak{J}(x)} \nonumber \\
\frac{\partial J}{\partial t}&=&\mathbb{R}\mathbb{H}(J,x)|_{x\,\,is\,\,a\,\,fixed\,\,parameter},
\end{eqnarray}
The symbol $\mathbb{R}\mathbb{H}$ denotes again an operator that is at this point undetermined,
Contrary to (\ref{zeta}) where the time evolution of $(x,J)$ is coupled, $x$ and $J$ evolve in (\ref{clhi}) separately. We assume that $J$ evolves faster than $x$. The time evolution of $J$ precedes the time evolution of $x$. For example, we can see (\ref{clhi}) as a closed Grad hierarchy.

All dynamical systems introduced above are assumed to be  well established and autonomous.  By "well established" we mean that predictions of the theory (solutions of its governing equations) agree with results of certain family of experimental observations. The family  is different for different levels. For example if (\ref{hyd10}) represents hydrodynamics then the family  consists of the observations that emerged  in the experimental hydrodynamics. Different observations are made for example if (\ref{hyd10}) represents kinetic theory (e.g. (\ref{hyd10}) represents the Boltzmann kinetic equation).
A theory is autonomous if it is well established and self contained.

In externally unforced systems $x$ in (\ref{hyd10}) and $(x,J)$ in (\ref{zeta}) approach  equilibrium states (fixed points in the state space). The approach is driven by  thermodynamic potentials that, if evaluated in the equilibrium states, become the fundamental thermodynamic relations in  equilibrium thermodynamics. These are well known results recalled in Section \ref{ThR1}.

There are two ways to see the relation  between (\ref{hyd10}) and (\ref{zeta}). In the first view  we regard (\ref{zeta}) as primary and (\ref{hyd10}) as its reduction. For example, (\ref{zeta}) is the kinetic equation (rewritten into the form of Grad hierarchy \cite{Grad}) and (\ref{hyd10}) are hydrodynamic equations. The second way to see the relation  between (\ref{hyd10}) and (\ref{zeta}) is to regard (\ref{hyd10}) as primary and (\ref{zeta}) as its extension.
There are three ways to make an extension. The first is a geometrical lift to higher order cotangent and tangent bundles discussed in \cite{OGM} and briefly sketched in Section \ref{ThR12}. The remaining two are physically motivated. The first by realizing that the internal structure that is not explicitly seen in (\ref{hyd10}) plays an important role in the time evolution and cannot be neglected. The second is motivated by realizing that it is  necessary  to include an extra inertia into the time evolution. The former follows  the  pioneering work of  Kirkwood \cite{Kirkwood1}, \cite{Kirkwood2} (that gave rise  to the modern theoretical rheology   \cite{Bird}) where the internal structure enters on the level of kinetic theory and another pioneering work of Cosserat brothers \cite{Cos} where the internal structure enters on the level of continuum theory \cite{Eck}, \cite{Van}, \cite{deLeon}. The second way of physically motivated extension
is inspired by  Grad's reformulation \cite{Grad}, \cite{GorbKar}, \cite{Str}) of the Boltzmann kinetic equation (that gave rise to extended thermodynamics  \cite{MullRug},\cite{RugAk} and irreversible thermodynamics \cite{Joubook}).

The vector fields in (\ref{hyd10}), (\ref{zeta}), (\ref{clhi}), which will appear in a more specific form in the course of our analysis,  are  all covectors transformed into  vectors by a geometrical structure. We shall use the following terminology: (a) \textit{gradient dynamics} if the covector $x^*$ is gradient of a potential $\Phi(x)$  and the geometrical structure is a positive definite operator $\Lambda$; $\frac{\partial x}{\partial t}=-\Lambda\frac{\partial \Phi}{\partial x}$; (b) \textit{generalized gradient dynamics} if the covector is a gradient of a potential $\Phi(x)$ and the geometrical structure is a potential $\Xi(x^*,x)$, called a structure potential;
$\frac{\partial x}{\partial t}=-\frac{\partial \Xi}{\partial x^*}|_{x^*=\frac{\partial \Phi}{\partial x}}$; (c) \textit{semigradient dynamics} is the covector $x^*$v is a general covector $F(x)$ (having for example the physical interpretation of a an external force) and
the geometrical structure is a potential $\Xi(x^*,x)$; $\frac{\partial x}{\partial t}=-\frac{\partial \Xi}{\partial x^*}|_{x^*=F(x)}$; (d) \textit{Hamiltonian dynamics}  if the covector is a gradient of a potential $\Phi(x)$ and the geometrical structure is a Poisson bivector $L$; $\frac{\partial x}{\partial t}=L\frac{\partial\Phi}{\partial x}$. The terminology is standard except for the term "semigradient dynamics" in (c) that is new. The Poisson bivector $L$ is defined in the text following Eq.(\ref{generic}). We note that (a) is a particular case of (b) if the structure potential is chosen to be a quadratic potential  $\Xi=\frac{1}{2}<x^*,\Lambda x^*>$, where $<,>$ is the pairing.

It is important to emphasize that the passage from (\ref{zeta}) to (\ref{clhi}) involves typically a very  hard analysis of solutions to (\ref{zeta}) that depends on the specificity of the vector field $\mathbb{E}\mathbb{H}$. In this paper we are only suggesting a physically (thermodynamically) meaningful framework into which results of such analysis can be placed. Typically, the main objective of the analysis is the first equation in (\ref{clhi}) representing the reduced dynamics.
Our main contribution is in putting into focus the second equation in (\ref{clhi}).

The main results  are the following.

(1)  The second equation in (\ref{clhi}) implies the rate thermodynamics. The rate thermodynamic potential is the Rayleighian known from the Onsager variational principle \cite{Ray}, \cite{Ons1}, \cite{OnsM}, \cite{Gy}, \cite{Doi}. The second equation in (\ref{clhi}) is the mathematical formulation of the zero law of rate thermodynamics similarly as Eq.(\ref{hyd10}) describing the time evolution in externally unforced systems (e.g. the Boltzmann kinetic equation) is the mathematical formulation of the zero law of thermodynamics.

(2) In externally unforced systems, the dissipative part of the vector field $\mathbb{H}$  in the first equation in (\ref{clhi}) is the generalized gradient dynamics with the structure potential that is the Legendre transformation of the rate entropy.
This means that in the case of the quadratic rate entropy  the rate of entropy equals (up to a multiplicative factor) the Legendre transformation of the rate entropy.

(3) In the context of externally driven systems, the vector field $\mathbb{H}$  in the first equation in (\ref{clhi}) is the semigradient dynamical systems with the Legendre transformation of the rate entropy as the structure potential. One of the consequences of this feature of (\ref{clhi}) is  that in the vicinity of critical points the dramatic increase of fluctuations translates  in the reduced dynamics governed by the first equation in (\ref{clhi}) into  the  appearance of a bifurcation.

(4) Some of the results of a general nature are illustrated on the example of mesoscopic thermodynamics and dynamics of the van der Waals gas.  A novel kinetic equation that is compatible with the van der Waals equilibrium thermodynamics is proposed.

\section{Externally unforced systems}\label{ThR1}

Externally unforced systems  are seen in experimental observations to approach, as $t\rightarrow\infty$, equilibrium sates  at which no time evolution takes place and their behavior is seen to be well described by equilibrium thermodynamics. This observation is called zero law of thermodynamics.  We investigate it in Section \ref{ThR11}.
In the more microscopic setting in which  $(x,J)$  serve as state variables the approach to equilibrium is gradual. First the state variable $J$ follows the time evolution governed by the second equation in (\ref{clhi}). This time evolution  brings $J$ to $\widehat{J}_{eq}(x)$. Subsequently the time evolution governed by the first equation in (\ref{clhi}) (in which $J=\widehat{J}_{eq}(x)$)  brings $x$ to the equilibrium value $x_{eq}$. The state variable  $J$ becomes  $J_{eq}=\widehat{J}_{eq}(x_{eq})$.
The approach $J\rightarrow \widehat{J}_{eq}$ gives rise to the rate thermodynamics (discussed in Section \ref{ThR12}).

\subsection{Classical dynamics and thermodynamics}\label{ThR11}

What is the structure of (\ref{hyd10}) that guarantees the experimentally observed approach to the equilibrium state $x_{eq}$? To answer this question we let ourselves to be guided by the Lyapunov theorem (the Lyapunov function, that  indicates  the approach to $x_{eq}$,  could be the thermodynamic potential) but mainly by the pioneering results of Boltzmann about dynamics of  ideal gases \cite{Boltzmann}. When dealing with externally unforced systems, Eq.(\ref{hyd10}) takes the form of the generalized gradient dynamics
\begin{equation}\label{eqXi}
\frac{\partial x}{\partial t}=-\frac{\partial \Xi(x^*,x)}{\partial x^*}|_{x^*=\frac{\partial \Phi}{\partial x}}
\end{equation}
The structure potential $\Xi(x^*,x)$ is called a dissipation potential and $\Phi(x)$ is the thermodynamic potential (both specified below). Particular realizations of (\ref{eqXi}) (i.e. (\ref{eqXi}) in which  $x$, $\Xi$ and $\Phi$ are specified)   are for instance  the Boltzmann kinetic equation without the free flow term \cite{Boltzmann}, \cite{book}, the Cahn Hilliard equation \cite{CH}, and the Ginzburg Landau equation \cite{GL}. The state variable $x$ is often a distribution function or a field. In such case the derivative with respect to $x$ is a functional derivative. In order to keep the notation simple, we use the same symbol $\frac{\partial}{\partial x}$ for both partial and functional derivatives.

 A more complete equation (GENERIC equation) that still guarantees approach to the equilibrium state will be introduced in Section \ref{ThR12}. Mathematically rigorous investigations of the approach to equilibrium reveal that the existence of the Lyapunov function is only an indication of the approach. They however reveal also   features of the approach of great physical significance. It has been shown in the context of the Boltzmann equation \cite{GradV}, \cite{Vill} that the free flow term on the right hand side of the Boltzmann equation, that by itself does not generate any dissipation (any change of the Lyapunov function) enhances the dissipation generated by the collision term. The local Maxwell distribution that would be the equilibrium state with the collision term alone, changes in the presence of the free flow term into the total Maxwell distribution. This mechanism (Grad-Villani-Desvillettes    enhancement of dissipation) is likely of fundamental importance in the onset of dissipation and the time irreversibility. We shall see some aspects of the Grad-Villani-Desvillettes     enhancement later in the analysis of the passage from (\ref{zeta}) to (\ref{clhi}).

The dissipation potential $\Xi$ is a real valued function of $(x^*,x)$ satisfying the following properties:
\begin{eqnarray}\label{Xiprop}
&&(i)\,\, \Xi|_{x^*=0}=0\nonumber \\
&& (ii)\,\, \Xi\,\,reaches\,\, its\,\, minimum\,\,  at\,\, x^*=0\nonumber \\
&&(iii)\,\, \Xi\,\, is\,\, convex\,\, in\,\, a\,\, neighborhood\,\, of\,\, x^*=0
\end{eqnarray}
Consequently, in a small neighborhood of $x^*=0$,  all dissipation potentials have the form $\Xi(x^*,x)=<x^*,\Lambda x^*>$, where $<,>$ is the pairing and $\Lambda$  a positive definite operator. In such case Eq.(\ref{eqXi}) becomes $\frac{\partial x}{\partial t}=-\Lambda\frac{\partial \Phi}{\partial x}$.

The properties of $\Xi$ listed in (\ref{Xiprop}) imply
\begin{equation}\label{Xiineq}
\frac{d\Phi}{dt}=-<x^*,\frac{\partial \Xi}{\partial x^*}>|_{x^*=\frac{\partial \Phi}{\partial x}}<0
\end{equation}
The thermodynamic potential $\Phi(x)$ thus indeed plays the role of the Lyapunov function (provided $\Phi(x)$ is a convex function) indicating  the approach of solutions to (\ref{eqXi}) to the equilibrium state $x_{eq}$ that is a solution to
\begin{equation}\label{EQPHI}
\frac{\partial \Phi(x)}{\partial x}=0
\end{equation}
The equilibrium states can be found either by following the time evolution governed by (\ref{eqXi}) to its conclusion or alternatively as states that minimize the thermodynamic potential (Maximum Entropy principle  MaxEnt).

Now we turn to  the thermodynamic potential
\begin{equation}\label{Phi}
\Phi(x,y^*))=-S(x)+<y^*,Y(x)>
\end{equation}
and to the equilibrium thermodynamics that it implies.
By $S(x)$ we denote  the  entropy, $Y(x)$ is the mapping $x\mapsto y=(E(x),N(x))$, where $E(x)$ is the energy and $N(x)$ number of moles, $y^*=(\frac{1}{T},-\frac{\mu}{T})$, $T$ is the temperature in energy units and $\mu$ is the chemical potential.
The equilibrium thermodynamics that is implied by  (\ref{hyd10}) is a collection of   its  two implications: (i)
equilibrium states $x_{eq}(y^*)$ approached in the time evolution are minima of the thermodynamic potential $\Phi(x)$ (MaxEnt),  and (ii) the equilibrium  fundamental thermodynamic relation in the state space of the classical equilibrium thermodynamics is
\begin{equation}\label{fthr1}
S^*(y^*)=\Phi(x_{eq}(y^*))
\end{equation}
where $S^*(y^*)$ is the Legendre transformation of the equilibrium entropy.

In order to complete the formulation of the classical thermodynamics, we still need to make  one step which relates  $S^*(y^*)$  to the therodynamic pressure $P$. So far we have not addressed the size of the system under investigation. We choose volume $V$ to characterize it and we adopt it  as an extra state variable. Next, we  extend  $S=S(E,N)$ (that is a Legendre transformation    of (\ref{fthr1}))  to $S=S(E,N,V)$. Moreover,  we assume that this extended function is one homogeneous in the sense that $\lambda S=S(\lambda E, \lambda N, \lambda V)$, holds for all  $\lambda\in \mathbb{R}$. The Euler relation $S=\frac{\partial S}{\partial E} E  + \frac{\partial S}{\partial N} N +\frac{\partial S}{\partial V} V $  together with $dS= \frac{\partial S} {\partial E} dE + \frac{\partial S}{\partial N} dN +  \frac{\partial S}{\partial V} dV$ and with $\frac{\partial S}{\partial E}= E^*=\frac{1}{T};\,
\frac{\partial S}{\partial N}= N^*=-\frac{\mu}{T};\,\frac{\partial S}{\partial V}= V^*=\frac{P}{T}$  then imply
\begin{equation}\label{Pre1}
S^*\left(\frac{1}{T},-\frac{\mu}{T}\right)=-\frac{P}{T}
\end{equation}
which is the complete equilibrium fundamental thermodynamic relation (i.e. fundamental thermodynamic relation on the level of the equilibrium thermodynamics) implied by the thermodynamic potential (\ref{Phi}).

Historically the first passage  from (\ref{Phi}) to (\ref{Pre1}) described above was made by Boltzmann \cite{Boltzmann} in his analysis of dynamics and thermodynamics of ideal gases. In this example $x$ is the one particle distribution function $f(\rr,\vv)$, Eq.(\ref{eqXi}) is the Boltzmann equation with the free flow term missing and with the cosh-dissipation potential $\Xi$  (see Eq.(4.84) in \cite{book}),  the entropy $S(f)$ is the Boltzmann entropy $-k_B \int d\rr\int d\vv f\ln f$ where $k_B$ is the Boltzmann constant,  energy $E(f)$ is the kinetic energy $\int d\rr\int d\vv  f \frac{\vv^2}{2m}$ where $m$ is the mass of one particle, and $N(f)=\int d\rr\int d\vv f$. The resulting relation (\ref{Pre1}) is the equilibrium fundamental thermodynamic relation  that in equilibrium thermodynamics represents ideal gases.

\subsection{Rate dynamics and thermodynamics}\label{ThR12}

Still considering only  externally unforced systems, we investigate the approach to equilibrium in the setting of a more microscopic dynamical theory in which $(x,J)$ serve as state variables. There are two ways to proceed. We can either follow the time evolution governed by (\ref{zeta}) or the time evolution governed  by its reformulation (\ref{clhi}). In the latter viewpoint the time evolution proceeds  in two stages. In the first (fast) stage $J \rightarrow \widehat{J}_{eq}(x)$ and in the second (slow) stage the passage $x\rightarrow x_{eq}$ is governed the first equation in (\ref{clhi}) with $J=\widehat{J}_{eq}(x)$.

In the  former viewpoint we can exactly repeat the analysis in Section \ref{ThR11}. On the level of abstraction on which we carried  the investigation of equilibrium thermodynamics in the previous section, the only difference between (\ref{hyd10}) and (\ref{zeta}) is that the state variable $x$ is replaced by $(x,J)$. In other words, we can simply see $x$ in the previous section as  $(x,J)$.

The latter viewpoint brings new thermodynamics. The approach  $J\rightarrow \widehat{J}_{eq}(x)$ provides the basis for the rate thermodynamics. We recall that we are assuming that all the dynamical systems introduced in Section \ref{Intr}  are well established. This means that the approach $J\rightarrow \widehat{J}_{eq}(x)$ must exist. We call its existence \textit{zero law of rate thermodynamics} similarly as the existence of the approach $(x,J)\rightarrow (x,J)_{eq}$ is called zero law of thermodynamics. The equilibrium state $(x,J)_{eq}$ reached by following the time evolution governed by (\ref{zeta}) is  $(x_{eq},\widehat{J}_{eq}(x_{eq}))$  if we see it as the equilibrium state reached by following the time evolution governed by (\ref{clhi})

 In the  formulation of the rate thermodynamics based on $J\rightarrow \widehat{J}_{eq}(x)$ we follow  the formulation of thermodynamics based on $(x,J)\rightarrow (x,J)_{eq}$ that we described in  Section \ref{ThR11}. The second equation in (\ref{clhi}) is assumed to be  the generalized gradient dynamics
(cf. (\ref{eqXi}))
\begin{equation}\label{REq}
\frac{\partial J}{\partial t}=\frac{\partial \Upsilon(J^{\dag},J,x)}{\partial J^{\dag}}|_{J^{\dag}=\frac{\partial \Psi(J)}{\partial J}}
\end{equation}
The structure potential $\Upsilon(J^{\dag},J,x)$, called  a rate dissipation potential, is required to   satisfy the three properties listed in (\ref{Xiprop}). By $\Psi(J)$ we denote rate dissipation potential that is defined below in (\ref{Psi}).
When discussing the rate time evolution of $J$ we use the symbol $^{\dag}$ to denote conjugate variable instead of the symbol $^*$ that we have been using  for the same purpose in the time evolution of $x$. When comparing (\ref{REq}) with (\ref{eqXi}) we note the difference in the sign of their right hand sides. The difference will be explained when a particular case
\begin{equation}\label{Ups2}
\frac{\partial J}{\partial t}=\mathbb{G}\frac{\partial \Psi(J)}{\partial J}
\end{equation}
 of (\ref{REq}), corresponding to quadratic rate dissipation potentials
\begin{equation}\label{Ups}
\Upsilon(J^{\dag},J)=\frac{1}{2}<J^{\dag},\mathbb{G} J^{\dag}>|_{J^{\dag}=\frac{\partial \Psi(J)}{\partial J}}
\end{equation}
where $\mathbb{G}$ is a positive definite operator,  will appear below as a cotangent lift of the GENERIC  extension  (\ref{generic}) of (\ref{eqXi}).
In the rest of this paper we  limit ourselves to the quadratic rate dissipation potentials and consider (\ref{Ups2}) as the time evolution equation representing the rate thermodynamics.

The rate thermodynamic potential
\begin{equation}\label{Psi}
\Psi(J,J^{\dag})=-\mathfrak{S}(J)+<J^{\dag},J>
\end{equation}
is a sum of  $-\mathfrak{S}(J)$, where $\mathfrak{S}$ is the rate entropy, and  $<J^{\dag},J>$, where $J^{\dag}$ are Lagrange multipliers.
Similarly as in the previous section, Eq.(\ref{Ups2}) implies
\begin{equation}\label{Upsin}
\frac{d\Psi}{dt}=<J^{\dag},\mathbb{G}J^{\dag}>|_{J^{\dag}=\frac{\partial \Psi}{\partial J}}>0
\end{equation}
Consequently, $-\Psi(J,J^{\dag})$ plays the role of the Lyapunov function (provided it is  a convex function function of $J$) and thus
the equilibrium state $\widehat{J}_{eq}(x)$, approached as $t\rightarrow\infty$, is $J$ at which $\Psi$ reaches it maximum  and thus $\widehat{J}_{eq}(x)$ is a solution to
\begin{equation}\label{Upsineq}
\frac{\partial \Psi(J,J^{\dag})}{\partial J}=0
\end{equation}
We also note that contrary to the situation in the classical thermodynamics where the entropy increases in the course of the time evolution, the rate entropy decrease in the course of the rate time evolution. The maximum entropy principle (MaxEnt) in thermodynamics  changes to the minimum rate entropy principle (MinRent) in the rate thermodynamics.

The  rate state variable in the second equation in (\ref{clhi}) does not have to be $J$ but it could be another state variable, we denote it $f$, through which the fluxes $J$ are expressed by $J=J(f)$.
For example (see Section \ref{vdW}), $f$ can be  one particle distribution function $f(\rr,\vv)$.  In such more general formulation of rate time evolution the rate thermodynamic potential $\Psi$ takes the form
\begin{equation}\label{Psig}
\Psi(f,J^{\dag})=-\mathfrak{S}(f)+<J^{\dag},J(f)>
\end{equation}
which then completely resembles (\ref{Phi}). If we see (\ref{Ups2})  in the context of (\ref{clhi}) then a   specification  $J^{\dag}=J^{\dag}(x)$  amounts to the the closure since the equilibrium state $\widehat{J}_{eq}$ reached as $t\rightarrow\infty$ is a solution to (\ref{Upsineq}). If we replace in (\ref{Ups2}) $J$ with $f$ and
$\Psi(J,J^{\dag})$  with  $\Psi(f,J^{\dag}(x))$ we obtain the equilibrium state $f_{eq}(J^{\dag}(x))$.

Similarly as in thermodynamics we have arrived at the fundamental relation of equilibrium thermodynamics we arrive now at  the fundamental relation of rate thermodynamics.
The rate thermodynamic potential  $\Psi$  evaluated at the rate equilibrium $f_{eq}(x)$  becomes  (cf. (\ref{fthr1}) the fundamental rate thermodynamic relation
\begin{equation}\label{fthr2}
\mathfrak{S}^{\dag}(J^{\dag}(x))=\Psi(f_{eq}(J^{\dag}(x)),J^{\dag}(x))
\end{equation}

At this point we again emphasize that the rate thermodynamics formulated above, if seen in the context of the dynamical theory (\ref{zeta}),  is just a framework into which results of an analysis of solutions to (\ref{zeta}) (a pattern recognition process in its phase portrait) can be placed. The framework has a clear physical meaning. It extends the classical thermodynamic  towards dynamics of externally driven systems. Moreover, the framework  provides  new connections in the network of various fields of nonequilibrium thermodynamics.
In particular, the rate thermodynamics connects the   Onsager variational principle, investigations of closures in dynamical hierarchies, and extended thermodynamics. Regarding the Onsager principle, Eq.(\ref{Ups2}) provides it with the time evolution and physical foundation.
We shall see this, as well as
connections to  closures in dynamical hierarchies and to  extended thermodynamics  later in this paper.

We now introduced more structure into (\ref{hyd10}) and (\ref{zeta}) that will allow us to enter more deeply into the passage (\ref{zeta}) $\rightarrow$ (\ref{clhi}). The specification  (\ref{eqXi}) of (\ref{hyd10}) applied to  externally unforced systems does provide a setting that unifies for instance the Navier Stokes Fourier  hydrodynamic equations without the Euler term with the Boltzmann equation without the free flow term. In order to find a setting  that unifies both equations in their complete form, we generalize (\ref{eqXi}) to
\begin{equation}\label{generic}
\frac{\partial x}{\partial t}=L T\frac{\partial \Phi}{\partial x}-\frac{\partial \Xi(x^*,x)}{\partial x^*}|_{x^*=\frac{\partial \Phi}{\partial x}}
\end{equation}
called a GENERIC equation \cite{GEN1},\cite{GEN11}, \cite{GEN12}, \cite{GEN13}, \cite{GEN14},  \cite{GEN2}, \cite{GEN3}, \cite{GEN4}, \cite{GEN5}, \cite{book}. In the first term on the right hand side, T is the temperature appearing in the thermodynamic potential $\Phi$ in (\ref{Phi}),
the operator $L$ is a Poisson bivector (i.e. $\{A,B\}=<\frac{\partial A}{\partial x},L
\frac{\partial A}{\partial x}> $ is a Poisson bracket;  $A$ and $B$ are real valued functions of $x$). The Poisson bracket has to satisfy the antisymmetry  $\{A,B\}=-\{B,A\}$ and the Jacobi identity $\{\{A,B\},C\}+\{\{B,C\},A\}+\{\{C,A\},B\} =0$  guaranteeing its conservation in the course of the Hamiltonian time evolution.
From the physical point of view, the Poisson bracket (or equivalently the Poisson bivector $L$) expresses mathematically the Hamiltonian kinematics of the state variable $x$.
In the particular case when $x=(\rr,\vv)$ then $L=\left(\begin{array}{cc}0&1\\-1&0\end{array}\right)$ is the canonical Poisson bivector.
Another specific  example  of $L$, expressing the Hamiltonian kinematics of the one particle distribution function,  will appear later in this paper (see (\ref{PBk})). A detail mathematical definition of the Poisson bivector as well as an explanation of its geometrical significance can be found for instance in \cite{OGM}.
The first term on the right hand side of (\ref{generic}), that is a new addition to the right hand side of (\ref{eqXi}), represents mesoscopic Hamiltonian dynamics that is an inheritance received by mesoscopic theories from the microscopic Hamiltonian dynamics of $\sim 10^{23}$ particles composing macroscopic systems. Since $<\frac{\partial\Phi}{\partial x},L\frac{\partial\Phi}{\partial x}>=0$ due to the antisymmetry of $L$, the inequality (\ref{Xiineq}) sill holds for (\ref{generic}) and thus the thermodynamic potential $\Phi$ plays the role of the Lyapunov function for the approach to the equilibrium state.

The GENERIC equation introduced in  \cite{GEN1},\cite{GEN11}, \cite{GEN12}, \cite{GEN13}, \cite{GEN14},   \cite{GEN3}, \cite{GEN4}, \cite{GEN5}, \cite{book} involves still more structure. The potential generating the Hamiltonian dynamics (i.e. the potential appearing in the first term on the right hand side of (\ref{generic})) is only the energy $E(x)$ and the potential generating the dissipative part (i.e. the potential appearing in the second term on the right hand side of (\ref{generic})) is only the entropy $S(x)$. This is achieved by requiring that $L$ is degenerate (in the sense that $L\frac{\partial S(x)}{\partial x}=0$, and $L\frac{\partial N(x)}{\partial x}=0)$  and the dissipation potential is degenerate in the sense that $\Xi$ depends on $x^*$ only through its dependence on $X=Kx^*$, where $K$ is a linear operator that satisfies $K\frac{\partial E(x)}{\partial x}=0, K\frac{\partial N(x)}{\partial x}=0$. For the sake of simplicity,  we do not require and do not use this extra structure in this paper. From the physical point of view, we are limiting
our investigation to isothermal systems.

We are now in position to show how the rate dynamic equation (\ref{Ups2}) arises as a cotangent lift of (\ref{generic}).  From the geometrical point of view, we want to lift the time evolution taking place in the state space to its cotangent space. Such geometrical lift is made in \cite{OGM}. In this paper we make the lift by  considering $x^*$ in (\ref{generic}) as an independent variable and rewriting (\ref{generic}) into  an equation governing the time evolution of $x^*$. We note that (\ref{generic}) with $x^*$ considered as an independent variable  can be rewritten as
\begin{equation}\label{genxstar}
\frac{\partial x}{\partial t}=\frac{\partial}{\partial x^*}\left(T<x^*,L\frac{\partial \Phi(x)}{\partial x}> -\Xi(x^*,x)\right)
\end{equation}
This equation will become  an equation governing the time evolution of $x^*$ if its left hand side is rewritten
in terms of $x^*$.
If $x^*= \frac{\partial \Phi}{\partial x}$ then  (\ref{genxstar}) becomes
\begin{equation}\label{genxstar1}
\frac{\partial x^*}{\partial t}=\widehat{\mathbb{G}}\frac{\partial}{\partial x^*}\left(T<x^*,L\frac{\partial \Phi(x)}{\partial x}> -\Xi(x^*,x)\right)
\end{equation}
where $\widehat{\mathbb{G}}=\frac{\partial^2 \Phi}{\partial x \partial x}$. We make  three observations about (\ref{genxstar}).

First, we note that if $x^*= \frac{\partial \Phi}{\partial x}$  then Eq.(\ref{genxstar}) is equivalent to (\ref{generic}). Equation (\ref{genxstar}) is an equivalent reformulation of (\ref{generic}) from the state variables $x$ to a new state variables $x^*$; the transformation $x\rightarrow x^*$ is one-to-one.

Second, we note that $\widehat{\mathbb{G}}=\frac{\partial^2 \Phi}{\partial x \partial x}$ is positive definite since $\Phi$ is a convex function. We now consider  $\widehat{\mathbb{G}}$ as an operator that is unrelated to $\Phi$ but remains still positive definite. With such interpretation, we denote it $\mathbb{G}$. Equation (\ref{genxstar}) with $\mathbb{G}$ replacing $\widehat{\mathbb{G}}$  is a new time evolution equation describing the time evolution in the cotangent space.

The momentum $J$ is an element of the cotangent space. If we identify it with $x^*$ then Eq.(\ref{Ups2}) is identical with (\ref{genxstar}). By comparing (\ref{genxstar}) with (\ref{Ups2}) we see that the dissipation potential  $\Xi$ plays the role of the rate entropy $\mathfrak{S}$ and also see the reason for the positive sign on the right hand side of (\ref{Ups2}).
The second term in the rate thermodynamic potential $\Psi$ in (\ref{Psi}) with $J^{\dag}=L\frac{\partial\Phi}{\partial x}$   is the rate of energy (rate of the free energy due to our limitation to isothermal systems). The thermodynamic potential $\Psi$ is thus the Rayleighian \cite{Ray} and (\ref{Ups2}) is the extension of the Onsager's variational principle to dynamics similarly as the Ginzburg Landau (\cite{GL}, Cahn Hilliard \cite{CH} or the GENERIC equation (\ref{generic}) extend MaxEnt to dynamics. We emphasize again that in both cases the physical basis of the extension to dynamics is the experimentally observed approach to reduced descriptions. In the former it is the  approach to a more macroscopic dynamics, in the latter it is the approach to equilibrium.

Summing up, Eq.(\ref{Ups2}) is a mathematical expression of the rate thermodynamics. It
provides a framework for placing  results of typically very extensive  analysis that is needed to pass from (\ref{zeta}) to (\ref{clhi}).  The results enter in the  rate entropy $\mathfrak{S}$ and the Lagrange multipliers $J^{\dag}(x)$ (both  appearing in the rate thermodynamic potential $\Psi$) and in the operator $\mathbb{G}$. The rate thermodynamic framework then gives
the closure and the fundamental relation of rate thermodynamics. The former (the flux  that maximizes the rate thermodynamic potential $\Psi$) gives the reduced dynamics (the first equation in (\ref{clhi})). The latter is an extra information about the structure of the reduced state space. We recall that the thermodynamic framework (Section \ref{ThR11}) also gives two information about the reduced state space (that is in this case the state space with elements $y=(E,N,V)$), namely the equilibrium state $x_{eq}(y)$ that lifts  the equilibrium state space (whose elements are $y$)  to a manifold inside a larger mesoscopic state space (whose elements are x) and
the  fundamental relation of equilibrium thermodynamics in the equilibrium state space.  In the rate thermodynamics it is the first result (i.e. the closure) that is most appreciated and most useful  while in thermodynamics the most appreciated is the second result (i.e. the fundamental thermodynamic relation). The role and the significance of the rate fundamental thermodynamic relation in the rate thermodynamics and the manifold structure of the thermodynamic state space  in equilibrium thermodynamics remain to be clarified.

\section{Externally driven systems}\label{ThR2}

Thermodynamics emerges from  investigations of     approaches
to  fixed points in  phase portraits of dynamical systems.
In externally unforced systems discussed in the previous section, we have seen two such approaches: the approach $x\rightarrow x_{eq}$ in the state space
and the approach $J\rightarrow \widehat{J}_{eq}(x)$ in the space  of vector field. The classical equilibrium thermodynamics is based on the former and the rate thermodynamics on the latter. In externally driven systems there is only one approach to fixed points, namely  $J\rightarrow \widehat{J}_{eq}(x)$. Consequently, the rate thermodynamics is the only thermodynamics  if the systems under investigations are subjected to external forces. The reduced dynamics (i.e. the time evolution of $x$) of externally driven systems is typically not driven by a potential \cite{HH}. If we look at  dynamics only in the  state space then,  in the presence of external forces, there is no thermodynamics.

The mathematical formulation of the rate thermodynamics of externally driven systems is the same as the formulation of the rate thermodynamics in externally unforced systems (presented in the previous section)  except that   the rate thermodynamic potential $\Psi$ (in particular then  the Lagrange multiplier $J^{\dag}(x)$) entering in it) is   different.  External forces enter the formulation in  the Lagrangian multipliers $J^{\dag}(x)$.

In order to clarify the exact meaning of the terms "fixed point" and "phase portrait"  we  note that all three dynamical systems (\ref{hyd10}), (\ref{zeta}), (\ref{clhi}) are in fact families of dynamical systems parametrized by quantities, $p \in \mathcal{P}$,  with which the individual nature of macroscopic systems is expressed. These quantities enter in the thermodynamic potentials as well as the geometrical structures transforming covectors into vectors.   The requirement that a dynamical system is well established specifies a subset in $\mathcal{P}$.

Phase portrait is by definition a collection of all trajectories (solutions of the time evolution equations). We use this concept in a more general sense. A phase portrait is a collection of all trajectories corresponding to the whole family of parameters $\mathcal{P}$ for which the dynamical systems are well established. Appearance of a fixed point in the phase portrait corresponding to a dynamical system with a single parameter $p\in \mathcal{P}$ does not give rise to thermodynamics.
Most important generalization of the search for attractive fixed points  that we are making  is that we are looking for attractive fixed points not only in phase portraits of dynamical systems but also in phase portraits of their extensions (lifts). In other words, our search for attractive fixed points and corresponding to them thermodynamics is multiscale.
While the emergence of thermodynamics from investigating fixed points in the phase portrait of dynamical systems is well known from Boltzmann \cite{Boltzmann}, the emergence of rate thermodynamics (applicable also to externally forced systems) from the same type of investigations in the phase portraits of extended (or lifted) dynamical systems is new.

Before continuing to discuss relations between thermodynamics and rate thermodynamics
we recall that there are phase portraits without attractive fixed points but  with thermodynamics. An important example is the phase portrait corresponding to the classical mechanics of $\sim 10^{23}$ particles composing macroscopic systems. In such situation it is necessary to use other    pattern  recognition methods  that are more sophisticated than just looking for attractive fixed points.
In fact, the search for fixed points in phase portraits  corresponding to lifted dynamics that we are making in this paper is one possible step towards such more sophisticated pattern recognition methods. Another method of this type can be based on results obtained by
Grad and Villani \cite{GradV}, \cite{Vill} in their investigation of solutions of the Boltzmann kinetic equation. A small nucleus of a microscopic instability grows  in the course of the time evolution into mesoscopic dissipation that then generates phase portraits with attractive fixed points.   Still another pattern recognition process in the microscopic phase portrait was introduced by Gibbs. In equilibrium statistical mechanics the pattern in the  phase portrait (the phase portrait corresponding to dynamics of $\sim 10^{23}$ particles) is searched as follows. First, the particle dynamics
is lifted (Liouville lift) to the space of real valued functions $f$ over the particle phase space (space of distribution functions). Second, the pattern emerges as a minimum of the thermodynamic potential (\ref{Phi}) in which the entropy $S(f)$ is universal for all macroscopic systems (Gibbs entropy) and the energy is the microscopic particle energy expressed as a function of $f$. From the physical point of view this means that the microscopic particle trajectories are assumed to be homogeneously spread (ergodic hypothesis). The energy and the number of particles (constants of motion)  are in fact the fixed points that  determine the pattern. In other words, the thermodynamics emerges from fixed points but the attraction to them is replaced by the ergodic hypothesis.

\section{Relation between thermodynamics and rate thermodynamics}\label{ThR14}

Equations (\ref{clhi}) (in their form (\ref{generic}), (\ref{REq}))  describing the time evolution of externally unforced systems involves two potentials,
$\Phi(x)$ and $\Psi(J)$. We may anticipate that they are related through their time evolution.  In this section we show that, roughly speaking, the  time derivative of $\Phi$ is the potential $\Psi$, or in other words that the rate entropy is essentially the  rate of entropy.  We emphasize that the two potentials  $\Phi(x)$ and $\Psi(J)$  are related only in externally unforced systems  simply because in externally driven systems the entropy $S(x)$ and the thermodynamic potential $\Phi(x)$ do not exist (since there is no approach to equilibrium).

For the purpose of this investigation we introduce into (\ref{zeta}) still more structure. We already know  that in order (\ref{zeta}) be compatible with thermodynamics in the sense of Section \ref{ThR1} it has to posses the GENERIC structure (\ref{generic}).
We now require that (\ref{zeta}) is a particular realization of (\ref{generic}) with a particular Poisson bivector $L$ and a particular dissipation potential
\begin{equation}\label{hier}
\frac{\partial}{\partial t}\left(\begin{array}{cc}x\\J\end{array}\right)=T\left(\begin{array}{cc}0&I\\-I&0\end{array}\right)
\left(\begin{array}{cc}\frac{\partial \Phi^{(ext)}}{\partial x}\\\frac{\partial \Phi^{(ext)}}{\partial J}\end{array}\right) -\left(\begin{array}{cc}0\\\frac{\partial \Theta}{\partial J^*}|_{J^*=\frac{\partial \Phi^{(ext)}}{\partial J}}\end{array}\right)
\end{equation}
The operator $I$ is the unit operator.
The matrix $L=\left(\begin{array}{cc}0&I\\-I&0\end{array}\right)$ is the canonical Poisson bivector.
By $\Phi^{ext}(x,J)$  we denote the thermodynamic potential that extends the thermodynamics potential $\Phi(x)$ to a more microscopic level on which $(x,J)$ serve as state variables.
The second term on the right hand side of (\ref{hier}) represents the dissipation, $\Theta$ is the dissipation potential that is required to satisfy the three properties listed in (\ref{Xiprop}). We are letting to dissipate explicitly only the extra state variable $J$ that addresses details not seen in $x$. From the physical point of view, it is indeed expected that the dissipation originates in instabilities in the time evolution that takes place on the  microscopic level. Such microscopic dissipation then trickles down to more macroscopic levels through its coupling in the Hamiltonian part (the first term on the right hand side of (\ref{hier})) of the mesoscopic dynamics. On the example of the Boltzmann equation this type on enhancement of dissipation was rigorously proven by Grad \cite{GradV}, and Villani, Desvillettes     \cite{Vill}.

We now reduce   (\ref{hier}) to (\ref{eqXi}). Such reduction needs a detailed analysis of solutions to (\ref{hier}).
As a first approximation,  we assume that the time evolution governed by the second equation in (\ref{hier}) with $x$ fixed drives the time evolution of $J$ to $J_{eq}(x^*)$.
The time evolution then continues as the time evolution of $x$ governed by the first equation in (\ref{hier}) with $J$ replaced by $J_{eq}(x)$. Direct calculations lead to  $0=-x^* - \frac{\partial \Theta}{\partial J^*}$
which, if solved for $J^*$,  becomes
\begin{equation}\label{int1}
J^*_{eq}(x^*)=\frac{\partial \Theta^{\dag}((J^*)^{\dag})}{\partial(J^*)^{\dag}}|_{(J^*)^{\dag}=-x^*}
\end{equation}
and to
\begin{equation}\label{Phislow}
\Phi(x)=\Phi^{(ext)}(x,J^*_{eq}(x))
\end{equation}
By $\Theta^{\dag}((J^*)^{\dag})$ we denote the Legendre transformation of $\Theta(J^*)$.
The first equation in (\ref{hier}) becomes
\begin{equation}\label{xx}
\frac{\partial x}{\partial t}=  J^*_{eq}(x) =       -\frac{\partial \Theta^{\dag}(x^*)}{\partial x^*}
\end{equation}
where
\begin{equation}\label{r1}
\Theta^{\dag}(x^*)= \Theta^{\dag}((J^*)^{\dag})|_{(J^*)^{\dag}=x^*}
\end{equation}
Identifying the dissipation potential  $\Theta(J^*)$ with the rate entropy $\mathfrak{S}(J^*)$  and also  $\Theta^{\dag}(x^*)$ with the Legardre transformation of the rate entropy $\mathfrak{S}^{\dag}(x^*)$
we arrive at
\begin{equation}\label{result}
\frac{d\Phi(x)}{d t}=-<x^*,\frac{\partial \mathfrak{S}^{\dag}(x^*)}{\partial x^*}>
\end{equation}
that relates the rate $\frac{d\Phi(x)}{dt}$  of the thermodynamic potential $\Phi(x)$ to  the rate entropy $\mathfrak{S}(x^*)$.

If we choose a quadratic dissipation  potential $\Theta(J^*)=\frac{1}{2}<J^*,\Lambda J^*>$, where $\Lambda$ is a symmetric and positive definite operator, then $\mathfrak{S}(x^*)=\frac{1}{2}<x^*,\Lambda^{-1}x^*>$, $\frac{\partial x}{\partial t}=\Lambda^{-1}x^*$,  and  $\frac{d\Phi(x)}{dt}=-<x^*,\Lambda^{-1}x^*>$. Consequently, in the case of the quadratic dissipation potential,  $\frac{d\Phi(x)}{dt}=-2\mathfrak{S}^{\dag}(x^*)$ (i.e.  the rate of the thermodynamic  potential equals mines two times the Legendre transformation of the rate entropy). In the case of a general dissipation potential satisfying (\ref{Xiprop}) the rate of the thermodynamic potential and  the rate entropy are related by (\ref{result}).

 We emphasize that in the above reduction from (\ref{hier}) to (\ref{xx}) we have passed  from one GENERIC equation (Eq.(\ref{hier})) to another GENERIC equation  (Eq.(\ref{xx})). In other words we are passing from one equation that is compatible with thermodynamics in the sense of Section \ref{ThR1} to another equation with the same property. We note that in order that the reduced equation (\ref{xx}) be GENERIC it is absolutely essential that the matrix $\left(\begin{array}{cc}0&I\\-I&0\end{array}\right)$  be skew symmetric. It is  this property that allows to transform  the gradient with respect to $J^*$ to gradient with respect to $x^*$.

\section{Critical behavior in thermodynamics and rate thermodynamics}\label{Cr}

The dynamical systems (\ref{hyd10}), (\ref{zeta}), and (\ref{clhi}), as they appear  in Section \ref{Intr}, have a  little   structure.
Investigations of
their compatibility with thermodynamics  in Sections \ref{ThR1}, \ref{ThR2}, and \ref{ThR14}, brought to them more  structure. In this section we bring still more structure and more physical insight into  the thermodynamic potentials entering them.

We begin with $\Phi(x)$ given in (\ref{Phi}).  The energy $E(x)$ and the number of moles $N(x)$ are mechanical concepts. Their specification requires knowledge of mechanical composition and interactions. In mechanics it is in the energy where the individual nature of systems is expressed. Shifting our view of macroscopic systems to mesoscopic views a new quantity, called entropy, joins the energy and the number of moles. How do we specify the entropy and what is its role in characterizing the individual nature of macroscopic systems?

The motivation for taking mesoscopic views is our interest in overall emerging features of solutions. The  role of the entropy is   to address this interest.
In  phase portraits the overall features appear as patterns. It is the role of the entropy to make the patterns  visible in the time evolution of appropriately modified microscopic dynamics.

This role of the entropy has first  emerged in  Boltzmann's analysis of  ideal gases. Boltzmann realized that binary collisions play the essential role in forming overall patterns in the phase portrait. In his analysis, made in the setting where one particle distribution function (and not the n-particle distribution function, $n\sim 10^{23}$) serves as the state variable, the free flow of particles remains exactly the same as in mechanics but binary collisions are treated as chemical reactions. Two particles enter the collision and two particles exit it, the mechanical details are ignored, only  the momentum and the kinetic energy are  required to be conserved. Boltzmann then arrived at the entropy (Boltzmann entropy) by analyzing solutions of the Boltzmann equation. In the context of our analysis, the Boltzmann investigation of solutions of Boltzmann's equation is  casting the Boltzmann equation into the form of (\ref{generic}) \cite{GEN1},\cite{book}.

In the absence of  the Boltzmann-like insight,  the entropy has to be specified by other means.
We have already discussed the Gibbs view of the equilibrium pattern in the phase portrait of the Liouville equation (that is a lift of the particle time evolution equations to the space of real valued functions over the particle state space), in Section \ref{ThR2}. In the Gibbs analysis the individual nature of systems is expressed in the energy (i.e. in the same way as in the microscopic mechanics). The entropy is universal for all systems. It only serves to sweep away  unimportant details in the phase portrait and reveal  the pattern.
As we have already mentioned in Section \ref{ThR2}, maximization of the Gibbs entropy  subjected to constraints (that are constants of motion,  i.e. energy and number of moles) reveals the equilibrium pattern.
As the viewpoint of macroscopic systems shifts to mesoscopic views, the entropy starts to take also the role of quantities expressing the individual nature of macroscopic  systems. Taking then the most macroscopic view, that is the view taken in equilibrium thermodynamics, it is only the  fundamental thermodynamic relation  where the individual nature is expressed. The fundamental thermodynamic relation is the entropy expressed as a function of the equilibrium thermodynamic state variables.   The energy in equilibrium thermodynamics serves as one of the equilibrium  state variable and is thus universal for all systems.  We  see that when  descending the scales from  completely microscopic to completely macroscopic, the role of entropy and the energy in characterizing the individual nature of systems changes. For the entropy, it increases from no role to remaining  the only quantity that plays this role. The completely opposite direction follow  the changes in the role of the energy.

Beside the universality of the  entropy in  Gibbs'  view of dynamics on the microscopic level, there is still another situation in which the thermodynamic potential (not only the entropy) is universal and the universality holds  not only on the microscopic level but on  all mesoscopic levels.
Since the entropy and thus also the thermodynamic potential are associated with patterns in  phase portraits,   we may expect that  in situations
when there is no pattern  the thermodynamic potentials generating the patterns  will be  essentially the same.
From the physical point of view, such  situations arise in critical points. These are the points when one pattern changes into another very different pattern (for example gas changes to liquid).
In the renormalization group theory of critical phenomena \cite{REN1}, \cite{REN2}, \cite{REN3} the critical points are in fact defined as fixed points of the renormalization procedure (that is in fact a pattern-recognition procedure, see more in Section \ref{vdW}). In a small vicinity of  critical points we may expect that the thermodynamic potentials will be universal. This is indeed the essential assumption on which the Landau theory of critical phenomena is based.

In the  context of multiscale thermodynamics the Landau theory \cite{Landau} is  the setting of Eq.(\ref{eqXi}) without the time evolution and restricted to the critical region. By (\ref{eqXi}) without the time evolution we mean that  the time evolution in (\ref{eqXi}) (that minimizes the thermodynamic potential $\Phi(x)$)  is replaced by postulating the minimization of $\Phi(x)$ (see more in Section \ref{vdW}).
The mesoscopic state variable $x$ is called in the Landau theory an order parameter. The critical point is the point at which determinant of the Hessian of $\Phi$ equals zero,  $det\frac{\partial^2 \Phi}{\partial x\partial x}=0$. The null space of the Hessian is assumed in the Landau theory to be one dimensional, $\xi\in\mathbb{R}$ are its coordinates.

At the critical point and in the null space of the Hessian  the thermodynamic potentials (considered to be smooth functions) take the form of critical  polynomials. Their classification and universality  is well studied and is known as catastrophe theory \cite{Thom}, \cite{Arnold2}.
The universality of critical polynomials  has been foreseen   by Landau and constitutes the basis of  his theory of phase transitions \cite{Landau}. We shall illustrate  mesoscopic thermodynamic potentials outside and inside the critical region  as well as  the time evolution associated with them in Section \ref{vdW} on the investigation of thermodynamics and dynamics of the  van der Waals gas. In the rest of this section we show how the criticality of  thermodynamic and rate thermodynamic potentials manifests itself  in the time evolution.

\subsection{Critical time evolution in externally unforced systems}\label{d20}

On the level on which $x$ serves as the state variable, the time evolution is governed by Eq.(\ref{eqXi}).
Jacobian $\mathfrak{J}$ of its  right hand side  at $x=x_{cr}$  is
\begin{equation}\label{Jacob}
\mathfrak{J}(x_{cr})=-\mathbb{J}(x_{cr})\mathcal{J}(x_{cr})
\end{equation}
where
\begin{equation}\label{Jacob1}
\mathbb{J}(x_{cr})=\left[\frac{\partial^2\Xi}{\partial x^*\partial x^*}|_{x^*=\frac{\partial\Phi}{\partial x}}\right]_{x=x_{cr}}
\end{equation}
and
\begin{equation}\label{Jacob2}
\mathcal{J}(x_{cr})=\left[\frac{\partial^2\Phi}{\partial x\partial x}\right]_{x=x_{cr}}
\end{equation}
Consequently, at the critical point,  $\det\mathfrak{J}(x_{cr})=0$  since  $\det\mathcal{J}(x_{cr})=0$.
This means that the appearance of a critical situation  appears in  observations of the approach to equilibrium as a bifurcation. There is, of course, a possibility that the dissipation potential $\Xi$  (not only the thermodynamic potential $\Phi$)   experiences critical behavior and consequently $\mathbb{J}$ in (\ref{Jacob}) contributes to its approach to zero as $x$ approaches $x_{cr}$. We shall see more about this point below.

Now we look at the approach to equilibrium in  the critical point from the point of view of the more microscopic theory on which $(x,J)$ serve as state variables and (\ref{zeta})  govern the time evolution.
As we have already noted,  the time evolution governed by (\ref{zeta}) can be seen in two ways, either as (\ref{hyd10}) considered in the previous paragraph  with  $x$  replaced by $(x,J)$ or as (\ref{clhi}). In the former case the investigation of the static and the dynamic critical behavior is the same as in the previous paragraph. The only difference is that   $x$ in (\ref{eqXi}) is replaced with $(x,J)$ and $\Phi(x,y^*)$ with $\Phi^{(ext)}((x,J),y^*)$.  The latter view in which the approach to equilibrium proceeds in two stages  (the fast time evolution of $J$ followed by the slower time evolution of $x$)  leads to a deeper insight into the dynamic and static  critical phenomena.

In Section \ref{ThR14} we have equipped  (\ref{zeta}) with the structure (\ref{hier}). We  continue with this particular realization of (\ref{zeta}). The time evolution of $J^*$ is governed by
\begin{equation}\label{d1}
\frac{\partial J^*}{\partial t}=\mathbb{G}\frac{\partial \Psi}{\partial J^*}
\end{equation}
where $\mathbb{G}$ is a positive definite operator and the rate thermodynamic potential
\begin{equation}\label{Psi2}
\Psi(J^*,x)=-\Theta(J^*)+<(J^*)^{\dag}(x^*), J^*>
\end{equation}
and
\begin{equation}\label{d2}
(J^*)^{\dag}(x^*)=-x^*
\end{equation}
The flux $J^*$ in (\ref{d1}) is considered to be an independent state variable.   In particular, we do not require  $J^*=\frac{\partial \Phi^{(ext)}}{\partial J}$.
If we follow the time evolution governed by (\ref{d1}) to its conclusion, $J^*$ reaches the state
(\ref{int1}). The subsequent time evolution of $x$ is then governed by (\ref{xx}).

We now ask the question as to whether the critical behavior seen in the behavior of $x$ is  also also seen in the behavior of the fluxes $J^*$ that address more microscopic details not seen in $x$. If we recall that the critical behavior is always accompanied with an increase of fluctuations and that larger fluctuations mean  larger role of microscopic details, the answer appears  to be affirmative. Similarly as
 the thermodynamic potential $\Phi(x)$ becomes critical at $x=x_{cr}$,  also the rate thermodynamic potential  $\Psi(J^*)$ becomes critical at $J^*_{cr}$. However, contrary to $\det \frac{\partial^2\Phi(x)}{\partial x\partial x}\rightarrow 0$ as $x\rightarrow x_{cr}$,
\begin{equation}\label{d3}
\det\frac{\partial^2\Psi(J^*)}{\partial J^*\partial J^*}\rightarrow \infty\,\, as\,\,J^*\rightarrow J^*_{cr}
\end{equation}
This is because of the dramatic increase of fluctuations when the critical point is approached and because the flux $J^*$ characterizes microscopic details that present themselves  in the experimental observations (made on the reduced level on which only $x$ serves as a state variable) as fluctuations.

Letting the fast evolution of $J^*$ to complete its course,   we turn to the next stage that is
the time evolution  of $x$  governed by (see (\ref{clhi}))
\begin{equation}\label{d7}
\frac{\partial x}{\partial t}=J^*
\end{equation}
With $J^*$ replaced by its value $J^*_{eq}$ reached in the fast time evolution, Eq.(\ref{d7}) becomes
\begin{equation}\label{d8}
\frac{\partial x}{\partial t}=\left[\frac{\partial \Theta^{\dag}((J^*)^{\dag})}{\partial (J^*)^{\dag}}|_{(J^*)^{\dag}=-x^*}\right]_{x^*=\frac{\partial\Phi}{\partial x}}=-\left[\frac{\partial\Theta^{\dag}(x^*)}{\partial x^*}\right]_{x^*=\frac{\partial\Phi}{\partial x}}
\end{equation}
where $\Theta^{\dag}((J^*)^{\dag})$ is the Legendre transformation of $\Theta(J^*)$ and \\$\Theta^{\dag}(x^*)=\Theta^{\dag}((J^*)^{\dag})|_{(J^*)^{\dag}=-x^*}$
The time evolution of $x$ is thus driven by gradient of a potential.
Jacobian $\mathfrak{J}$ of the right hand side of (\ref{d8}) is (\ref{Jacob}) with
\begin{equation}\label{d5}
\mathbb{J}=\left[\frac{\partial ^2 \Theta^{\dag}((J^*)^{\dag})}{\partial (J^*)^{\dag}\partial (J^*)^{\dag}}|_{(J^*)^{\dag}=x^*}\right]_{x=x_{cr}}
\end{equation}
and $\mathcal{J}$ is given in (\ref{Jacob2}).
Noting that  (\ref{d3}) implies  $\Theta^{\dag}((J^*)^{\dag})\rightarrow 0\,\, as \,\,x\rightarrow x_{cr}$
we conclude that $\det\mathbb{J}\rightarrow 0\,\,at\,\,x=x_{cr}$. Moreover,
in externally unforced systems  $(J^*)^{\dag}(x^*)=-x^*$ and thus also  $\det\mathcal{J}\rightarrow 0\,\,at\,\,x=x_{cr}$. Consequently,
$\det\mathfrak{J}\rightarrow 0\,\,at\,\,x=x_{cr}$ which means that the  critical behavior manifests itself in the approach to equilibrium as appearance of a bifurcation. We have thus arrived at the same result as when we considered the time evolution only on the level on which $x$ serves as the state variable (i.e. when the time evolution is governed by (\ref{eqXi})). The more microscopic viewpoint taken by adopting the flux $J^*$ as an extra state variable gave us the  dissipation potential $\Xi$ expressed in terms of the dissipation potential $\Theta$ appearing in the time evolution of $J^*$   (namely $\Xi=\Theta^{\dag}$ ) and thus another reason for  appearance of the bifurcation. The Jacobian $\mathfrak{J}$ tends to zero not only because the criticality appearing in the observations of $x$ (i.e. in $\det\mathcal{J}\rightarrow 0\,\,at\,\,x=x_{cr}$) but also because of the increase of fluctuations seen on the more microscopic level  (that is mathematically expressed in $\det\mathfrak{J}\rightarrow 0\,\,as\,\,x\rightarrow x_{cr}$).

\subsection{Critical time evolution in externally driven systems}\label{d21}

We have seen above that the time evolution in externally unforced systems is driven by gradient of a potential, namely by  gradient of the thermodynamic potential. This then makes a direct connection between the thermodynamic and dynamic critical behavior. The situation is different in the case of the externally driven systems.
It is well known \cite{HH} that the time evolution in  externally driven systems is not (at least typically) driven by  gradient of a potential. We shall show that the potential that drives the fluctuations (characterized by the flux $J^*$) participates in the time evolution of $x$ in a way that its critical behavior (dramatic increase of fluctuations) implies the critical behavior in the time evolution of $x$.

We assume that the external forces will enter our formulation in the time evolution of $J^*$. From the physical point of view, we assume that the external forces act directly only on the more microscopic state variable $J^*$. Their influence on  time evolution of $x$) is only indirect through the coupling in the time evolution of $J^*$ and $x$.
The time evolution of the flux $J^*$ is in the presence of external forces still governed by (\ref{d1}) with the rate thermodynamic potential (\ref{Psi2}) but with
\begin{equation}\label{d3}
(J^*)^{\dag}=F(x)
\end{equation}
replacing (\ref{d2}). By $F(x)$ we denote the external forces.
With this change,  $J^*$ reached as $t\rightarrow\infty$ is
\begin{equation}\label{JexF}
J^*_{eq}=\frac{\partial\Theta^{\dag}((J^*)^{\dag})}{\partial (J^*)^{\dag}}|_{(J^*)^{\dag}=F(x)}
\end{equation}
With $J^*$ in (\ref{d7}) replaced by (\ref{JexF}), we arrive at
\begin{equation}\label{d77}
\frac{\partial x}{\partial t}=\left[\frac{\partial \Theta^{\dag}((J^*)^{\dag})}{\partial (J^*)^{\dag}}|_{(J^*)^{\dag}=F(x)}\right]
\end{equation}
governing the time evolution of $x$. According to the terminology that we have introduced in Introduction, the dynamics with (\ref{d77}) as the time evolution equations id a semigradient dynamics.
Equation (\ref{d77}) cannot be cast to the form of (\ref{eqXi}) for all external forces $F(x)$. This means that the time evolution of externally driven systems on the mesoscopic level on which $x$ serves as the state variable are not driven (at least typically) by gradient of a potential. This result is in agreement with well known observations made about dynamics of externally driven systems \cite{HH}.

We turn now to the behavior in critical situations. We define the critical point on the level on which the fluxes $J^*$ play the role of state variables. In the rate thermodynamics we define the critical point $x_{xr}$ as a point on which
\begin{equation}\label{F1}
\det\frac{\partial^2 \Theta}{\partial J^*\partial J^*}|_{(J^*)^{\dag}=F(x)}\rightarrow \infty \,\,as\,\,x\rightarrow x_{cr}
\end{equation}
From the physical point of view, we define the critical point by the behavior of fluctuations (characterized mathematically by the fluxes $J^*$). We prove now that the critical behavior seen in fluctuations implies the appearance of a bifurcation in solutions of (\ref{d77}) which is the manifestation of the critical behavior of  $x$. Indeed, (\ref{F1}) implies
\begin{equation}\label{F2}
\det\mathbb{J}=\det\frac{\partial^2 \Theta^{\dag}}{\partial (J^*)^{\dag}\partial (J^*)^{\dag}}|_{(J^*)^{\dag}=F(x)}\rightarrow  0 \,\,as\,\,x\rightarrow x_{cr}
\end{equation}
which then  means that
\begin{equation}\label{F4}
\det\mathfrak{J}=\det\mathbb{J}|_{(J^*)^{\dag}=F(x)}{\det\frac{\partial F(x)}{\partial x}}\rightarrow  0 \,\,as\,\,x\rightarrow x_{cr}
\end{equation}
irrespectively of the behavior of $\det\frac{\partial F(x)}{\partial x}$ in the vicinity of the critical point $x_{cr}$.

Looking at macroscopic systems on a mesoscopic level $\mathcal{M}$ on which $x\in M$ serves  as state variable ($M$ denotes the mesoscopic state space), we have shown that if external forces act directly only on the  vector fields $J^*\in TM$, (or alternatively $J^*$ can have  a physical interpretation of  more microscopic details that manifest themselves only as fluctuations) then their dramatic increase in critical points is seen in the time evolution of $x$ as a bifurcation. The time evolution of $x$ in the presence of external forces is not driven by gradient of a  potential. The time evolution of $J^*$, in both the absence and the  presence of external forces,  is driven by a potential $\Psi(J^*)$. This is because  the mesoscopic level $\mathcal{M}$ is assumed to be well established (both in the absence and the  presence of external forces) and consequently $J^*$ has to become in the course of the time evolution  enslaved to the time evolution of $x$.
We have shown that the critical behavior of $\Psi(J^*)$ (seen on the mesoscopic level $\mathcal{M}$ as a dramatic increase of fluctuations) makes its appearance in the time evolution of $x$ as an appearance of a bifurcation.

This result has one common feature with results obtained in investigations of Landau damping \cite{GrPav} and Grad-Villani-Desvillettes  enhancement of dissipation \cite{GradV}, \cite{Vill}. In all three cases the Hamiltonian  coupling between mesoscopic and more microscopic time evolution) passes an instability  on the more microscopic level down to the more macroscopic levels. In the case of the Grad  Villani enhancement it is the passage of dissipation in the velocity variable  of  the one particle distribution function to the Navier-Stokes-Fourier dissipation in hydrodynamics. In Landau damping it is the passage of micro turbulence in velocities to the dissipation in the position coordinate. In this paper it is the passage from a critical behavior of $J^*$ (interpreted for example as an increase of fluctuations) to  a critical behavior on the mesoscopic level $\mathcal{M}$. The common mechanism for passing dissipation and instabilities down to more macroscopic levels plays likely an important role in the onset of dissipation and the time irreversibility in mesoscopic dynamics.

\section{Dynamics of the van der Waals fluid}\label{vdW}

Three dynamical models of the van der Waals gas (vdW gas) become in this section  three  particular realizations of the
dynamical systems (\ref{hyd10}), (\ref{zeta}), and (\ref{clhi}) introduced in Section \ref{Intr}. From the microscopic point of view, the particles composing the vdW gas differ from those composing the ideal gas. The ideal gas particles are point particles that do not interact among themselves. The vdW particles are finite size particles that interact among themselves via a hard core repulsion and long range attraction.  From the macroscopic point view,  the vdW gas exhibits the phase change from gas to liquid and the ideal gas does not.

We begin by recalling a mesoscopic  equilibrium thermodynamics of the vdW gas in Section \ref{vdW1}. In Section \ref{vdW2} we extend it to a mesoscopic dynamical theory and formulate particular realizations of (\ref{hyd10}), (\ref{zeta}), and (\ref{clhi}). For completeness, we  recall in  Section \ref{vdW3} the Enskog Vlasov kinetic theory of the vdW gas with a  particular attention to  problems with its link  to the equilibrium theory.

\subsection{Mesoscopic equilibrium theory}\label{vdW1}

We begin the formulation of mesoscopic models of the vdW gas on the level on which  the number of moles $n(\rr)$ serves as the state variable (i.e. $x=n(\rr); \rr\in \mathbb{R}^3$ is the position vector).   Our first task is to formulate the thermodynamic potential $\Phi(n(\rr),y^*)$ and then  investigate the critical behavior that it implies. The time evolution in the form of Eqs.(\ref{hyd10}),(\ref{zeta}), (\ref{clhi}) is considered in the following two sections.

The thermodynamic potential $\Phi(n(\rr))$ has been formulated in \cite{vKamp}, the critical phenomena that it implies in \cite{REN2}, \cite{REN3}. Here we only recall the  points that illustrate our investigation and that are needed to investigate the time evolution in the next two sections.

Following van Kampen \cite{vKamp}, the energy $E(n)$, the entropy $S(n)$, and the number of moles $N(n)$ are chosen as follows.
\begin{eqnarray}\label{vdWn}
S(n)&=&-k_B\int d\rr(n(\rr)\ln n(\rr)+k_Bn(\rr)\ln(1-bn(\rr)))\nonumber \\
E(n)&=&-\frac{3}{2}T\int d\rr n(\rr)+\frac{1}{2}\int d\rr \int d\rr_1  n(\rr)V_{pot}(|\rr-\rr_1|)n(\rr_1)\nonumber \\
N(n)&=&\int d\rr n(\rr)
\end{eqnarray}
The energy is the sum of the kinetic energy and the Vlasov mean field expression for the attractive potential energy, $V_{pot}(|\rr-\rr_1|)$ is the attractive potential. The expression for the number of moles $N(n)$ is a direct consequence of the physical interpretation of the state variable $n(\rr)$.

The entropy $S(n)$ is a sum of the Boltzmann entropy and the entropy expressing the constraint in the motion due to the hard core repulsive potential. The parameter $b$ is the diameter of the hard core. Here we see an example where, by replacing a mechanical force with a constraint  on  the motion, the entropy ceases to be  universal. It becomes  in   mesoscopic theories  one  of the  quantities in which   the individual nature of systems under investigation is expressed. The second  term in the entropy can alternatively be obtained \cite{book} by requiring that MaxEnt reduction to the equilibrium described in Section \ref{ThR1}) leads to the phenomenological van der Waals pressure-volume-temperature relation. The second  term can also be supplemented by a weakly nonlocal term $\widehat{b}\int d\rr n(\rr)\frac{\partial n(\rr)}{\partial \rr}\frac{\partial n(\rr)}{\partial \rr}$, where $\widehat{b}$ is a new  parameter. This nonlocal term does not change the implied fundamental thermodynamic relation but it plays a role in dynamics discussed in the next section.
Regarding the consideration of some mechanical forces as entropic forces, we
may  recall  the elasticity theory where  difference manifests itself as a difference between  the metal and rubber elasticity. We shall see also the consequences of the entropic view of the hard core force  in the next section in the investigation of dynamics.

Having the vdW thermodynamic potential, the following development of the vdW equilibrium theory is  routine (see details in \cite{vKamp}, \cite{book}, \cite{REN2},\cite{REN3}). The MaxEnt passage to the equilibrium leads to the classical van der Waals equilibrium thermodynamics. In the investigation of the critical behavior, we assume that $n(\rr)=n$ is independent of $\rr$ and see it as an order parameter in the setting of the Landau theory. The vdW thermodynamic potential implies (see details in \cite{book}) that the critical point is
\begin{eqnarray}\label{crp}
&&n_{cr}=\frac{1}{3b};\,\, T_{cr}=\frac{4V_{pot}}{27b}\nonumber \\
 && \left(\frac{\mu_{cr}}{T_{cr}}\right)=\frac{1}{2}\ln (3b)+\frac{3}{2}\ln\left(\frac{b}{V_{pot}}\right)+\frac{3}{4}+4\ln\frac{3}{4}-\frac{3}{2}\ln(2\pi)
\end{eqnarray}
and the Landau critical polynomial takes the form
\begin{equation}\label{Lpol}
\Phi^{(cr)}(\xi)=\alpha\xi+\frac{1}{2}\beta\xi^2+\frac{1}{24}\gamma \xi^4
\end{equation}
where $\xi=(n-n_{cr})$,
\begin{eqnarray}\label{Lancrexp}
\gamma &=&\frac{(27)^2b^3}{8}\nonumber \\
\beta&=&\beta_0\left(\frac{1}{T}-\frac{1}{T_{cr}}\right)\nonumber \\
\alpha &=&\alpha_0\left(\frac{\mu}{T}-\left(\frac{\mu}{T}\right)_{cr}\right)+\alpha_1\left(\frac{1}{T}-\frac{1}{T_{cr}}\right)
\end{eqnarray}
and $\alpha_0,\alpha_1$ and $\alpha_1$ are expressed in terms of $b,V_{pot},T,\mu$. The critical exponents implied by (\ref{Lancrexp}) are the classical exponents.

The mesoscopic van der Waals theory presented above demonstrates clearly  differences and common features of mesoscopic thermodynamics and the Landau theory of critical phenomena. The critical polynomial  (\ref{Lpol}), (\ref{Lancrexp}) is the starting point of the Landau theory.
The point of departure of
the mesoscopic thermodynamics is the thermodynamic potential (\ref{vdWn}).  The Landau polynomial emerges as one of its results. In addition, the mesoscopic thermodynamics  locates the critical point and provides a complete knowledge of the thermodynamic behavior both outside and inside the critical region.

In order to get the van der Waals critical exponents closer to  results of  experimentally observations, the van der Waals mesoscopic theory has to be  in the critical region renormalized.
In the context of the multiscale thermodynamics
the renormalization process introduced in the renormalization group theory of critical phenomena \cite{REN1} appears as  a pattern recognition process converging to the critical point that is defined as a point at which  no pattern can be recognized.  Let the static or the dynamic theory be initially formulated on a level $\mathcal{L}$, the parameters in which the individual nature of macroscopic systems  are expressed are $\mathcal{P}$. The first step is an extension to a more microscopic level $\mathfrak{L}$ with the parameters $\mathfrak{P}$. The second step is a reduction $\mathfrak{L}\rightarrow \mathcal{L}$ in which $\mathfrak{P}\rightarrow \mathcal{P}_1=\mathcal{P}_1(\mathfrak{P})$. This process continues until the extension followed by the reduction does not change the system. In the classical realization of the renormalization procedure the level $\mathcal{L}$ is the most microscopic level on which macroscopic systems are seen as composed of $\sim 10^{23}$ particles. The theory is the Gibbs equilibrium  statistical mechanics. Since there is no more microscopic level than the level $\mathcal{L}$, the level $\mathfrak{L}$ remains the same but the macroscopic system under investigation is enlarge in a sense that one particle composing it is replaced by a "box" of particles. The box is subsequently reduced to a "quasiparticle". In the critical point the quasiparticle is identical with the initial particle. Another realization of the renormalization process is the renormalization of the van der Waals theory that has been worked out in \cite{REN1},\cite{REN2}. The level $\mathcal{L}$ is the level on which we have formulated mesoscopic  thermodynamics of the vdW gas, the parameters $\mathcal{P}$ are the coefficients in the Landau critical polynomials. The more microscopic level is the level on which the state variable $n(\rr)$ becomes $(n(\rr),m(\rr))$. Initially one component system is seen as a two component system, $m(\rr)$ denotes the number of moles of the second component. The two components are identical, only the viewpoint changes. In the second step the extended formulation is reduced by MaxEnt to the initial one component view and rescaled to the initial number of moles.

\subsection{Extension  to dynamics}\label{vdW2}

On the level on which  $x=n(\rr)$ serves as the state variable, the time evolution is governed by
Eq.(\ref{eqXi})
\begin{equation}\label{eqXi1}
\frac{\partial n(\rr)}{\partial t}=-\frac{\partial \Xi(n^*,n)}{\partial n^*}|_{n^*(\rr)=\frac{\partial\Phi(n)}{\partial n}}
\end{equation}
with the thermodynamic potential given in (\ref{vdWn}). The dissipation potential $\Xi$ is an extra quantity that enters in the dynamical extension. The first natural choice is the quadratic dissipation potential $\Xi(n^*,n)=\int dx <n^*,\Lambda(n) n^*>$ in which the extra quantity entering dynamics is the positive definite operator $\Lambda(n)$.

The extension (\ref{eqXi1}) of the vdW mesoscopic thermodynamics to dynamics follows the route paved by Ginzburg Landau \cite{GL} and Cahn  Hilliard \cite{CH}. With the extension (\ref{eqXi1}) we have replaced the postulated MaxEnt (minimization of the thermodynamic potential $\Phi$)    with  Eq.(\ref{eqXi1})  that minimizes  $\Phi$ by following the time evolution that it generates (Eq.(\ref{eqXi1}) expresses mathematically the zero law of thermodynamics).

Now we turn to a more microscopic viewpoint in which the one particle distribution function $f(\rr,\vv)$ serves as the state variable. We construct the kinetic equation as a particular realization of the GENERIC equation (\ref{generic}). First, we extend the thermodynamics potential (\ref{vdWn}) to the kinetic theory level
\begin{eqnarray}\label{vdWf}
S(f)&=&-k_B\int d\rr\int d\vv(f(\rr,\vv)\ln f(\rr,\vv)+k_Bf(\rr,\vv)\ln(1-bn(\rr)))\nonumber \\
E(f)&=&\int d\rr \int d\vv \frac{\vv^2}{2}f(\rr,\vv)\nonumber \\
&&+\frac{1}{2}\int d\rr \int d\rr_1 \int d\vv \int d\vv_1  f(\rr,\vv)V_{pot}(|\rr-\rr_1|)f(\rr_1,\vv_1)\nonumber \\
N(f)&=&\int d\rr \int d\vv  f(\rr,\vv)
\end{eqnarray}
As we have already mentioned in (\ref{vdWn}), the entropy $S(f)$ can be extended by adding the  weakly nonlocal term $\widehat{b}\int d\rr\int d\vv f(\rr,\vv)\frac{\partial n(\rr)}{\partial \rr}\frac{\partial n(\rr)}{\partial \rr}$. We note that the MaxEnt reduction of the thermodynamical potential (\ref{vdWf}) leads to the Maxwell velocity distribution and the classical vdW equilibrium fundamental thermodynamic relation. In other words, the equilibrium fundamental thermodynamic relation implied by (\ref{vdWf}) is the same as the equilibrium fundamental thermodynamic relation implied by (\ref{vdWn}).

The next step in constructing a particular  realization of (\ref{generic}) is to specify the Poisson bivector $L$ expressing mathematically the kinematics of $f(\rr,\vv)$. The arguments recalled for example in \cite{book} lead to the Poisson bracket
\begin{equation}\label{PBk}
\{A,B\}=\int d\rr\int d\vv f \left(\frac{\partial \frac{\partial A}{\partial f}}{\partial \rr} \frac{\partial \frac{\partial B}{\partial f}}{\partial \vv} - \frac{\partial \frac{\partial B}{\partial f}}{\partial \rr} \frac{\partial \frac{\partial A}{\partial f}}{\partial \vv}\right)
\end{equation}
expressing kinematics of the one particle distribution function. With (\ref{BBk}), the GENERIC equation (\ref{generic}) becomes
\begin{equation}\label{vdWkeq}
\frac{\partial f(\rr,\vv)}{\partial t}=-\frac{\partial}{\partial \rr}\left(T f\frac{\frac{\partial\Phi}{\partial f}}{\partial \vv}\right) +\frac{\partial}{\partial \vv}\left(Tf\frac{\frac{\partial\Phi}{\partial f}}{\partial \rr}\right) -\frac{\partial\Xi}{\partial f^*(\rr)}|_{f^*(\rr)=\frac{\partial\Phi}{\partial f(\rr)}}
\end{equation}
Using the thermodynamic potential (\ref{vdWf}), the kinetic equation (\ref{vdWkeq}) gets the form
\begin{eqnarray}\label{Geq}
\frac{\partial f(\rr,\vv)}{\partial t}&=&-\frac{\partial}{\partial \rr}\left(\frac{\vv}{m}f(\rr,\vv)\right)\nonumber \\ &&+T\frac{\partial f(\rr,\vv)}{\partial \vv}\frac{\partial}{\partial \rr}\left(\ln (1-bn(\rr))-
\frac{bn(\rr)}{1-b\int dn(\rr))}\right)\nonumber \\
&&+\frac{\partial f(\rr,\vv)}{\partial\vv}\frac{\partial}{\partial\rr}\int d\rr_1 V_{pot}(|\rr-\rr_1|)n(\rr_1)\nonumber \\
&&-\frac{\partial\Xi(f,f^*)}{\partial f^*(\rr,\vv)}|_{f^*=\frac{\partial\Phi}{\partial f}}
\end{eqnarray}
where $n(\rr)=\int d\vv f(\rr,\vv)$.
We leave still the dissipation potential $\Xi$ undetermined. The most obvious choice is the Boltzmann dissipation potential which makes the last term on the right hand side of (\ref{Geq}) the Boltzmann collision term. From the physical point of view, this  means that the kinetic equation (\ref{Geq}) takes the hard core repulsive potential into account in the entropic term (the second term on its right hand side) and the collisions remains the same as in the ideal gas. Another possible choice is the symmetric part of the Enskog collision operator introduced in the next section.

If we include in  the entropy $S(f)$   the nonlocal term
$\widehat{b}\int d\rr\int d\vv f(\rr,\vv)\frac{\partial n(\rr)}{\partial \rr}\frac{\partial n(\rr)}{\partial \rr}$  then (\ref{Geq}) becomes
\begin{eqnarray}\label{Geq1}
\frac{\partial f(\rr,\vv)}{\partial t}&=&-\frac{\partial}{\partial \rr}\left(\frac{\vv}{m}f(\rr,\vv)\right)\nonumber \\ &&+T\frac{\partial f(\rr,\vv)}{\partial \vv}\frac{\partial}{\partial \rr}\left(\ln (1-bn(\rr))-
\frac{bn(\rr)}{1-bn(\rr))}\right)\nonumber \\
&&-T\frac{\partial f(\rr,\vv)}{\partial \vv}\frac{\partial}{\partial \rr}\left(\frac{\partial n(\rr)}{\partial \rr}\frac{\partial n(\rr)}{\partial \rr}+2n(\rr)\triangle n(\rr)\right)\nonumber \\
&&+\frac{\partial f(\rr,\vv)}{\partial\vv}\frac{\partial}{\partial\rr}\int d\rr_1 V_{pot}(|\rr-\rr_1|)n(\rr_1)\nonumber \\
&&-\frac{\partial\Xi(f,f^*)}{\partial f^*(\rr,\vv)}|_{f^*=\frac{\partial\Phi}{\partial f}}
\end{eqnarray}
where $n(\rr)=\int d\vv f(\rr,\vv) $ and $\triangle$ is  Laplacian.

Equations (\ref{Geq}) and (\ref{Geq1}) are novel kinetic equations addressing the time evolution  of the vdW gas. An investigation of their solutions as well as their comparison with other existing kinetic equations is expected to contribute to the understanding of dynamics of gas liquid phase transitions. We limit ourselves in this paper only to a few comments. Solutions of (\ref{Geq}) are discussed in the rest of this section and the comparison with the Enskog Vlasov kinetic equation in the next section.

Kinetic equations (\ref{Geq}) and (\ref{Geq1}) are
particular realizations of the GENERIC equation (\ref{generic}) and thus the inequality (\ref{Xiineq}) holds and the thermodynamic potential plays the role of the Lyapunov function for the approach to the equilibrium state that minimizes it. The thermodynamic potential entering
(\ref{Geq}) and (\ref{Geq1}) guarantees that at the approached equilibrium states the gas behaves as the vdW gas. This important property of solutions of  (\ref{Geq}) and (\ref{Geq1}),  guaranteeing  their  compatibility with the vdW equilibrium thermodynamics,  remains unproven for example for the Enskog Vlasov kinetic equation discussed in the next section. Regarding other properties of solutions, we outline  below a path that
follows formulations (\ref{zeta}), (\ref{clhi}).

Our  task is to recast (\ref{Geq}) (or (\ref{Geq1})  to the form (\ref{zeta}) and (\ref{clhi}). This requires an extensive analysis of solutions to (\ref{Geq}) and (\ref{Geq1}). We only indicate one possible route that may be taken to achieve it.   We begin by taking two completely independent views of the vdW gas. In the first view we see the gas on the level on which the number of moles $\nu(\rr)$ serves as the state variable.
We use the symbol $\nu(\rr)$ instead of $n(\rr)$ used in the previous analysis in order to keep $n(\rr)$ for a new combined number of moles introduced (see (\ref{trnnu}) below) in the second step.
The time evolution of $\nu(\rr)$ is governed by (\ref{eqXi1}) (with $n(\rr)$ replaced by $\nu(\rr)$). In the second view we see the gas on the level on which $f(\rr,\vv)$ serves as the state variable and (\ref{vdWkeq}) governs the time evolution. In this "double vision" viewpoint, the state variables are $(\nu(\rr), f(\rr,\vv))$ and the kinematics is mathematically expressed in the Poisson bracket (\ref{PBk}) since $\nu(\rr)$ has no Poisson kinematics. This double-vision  point of departure is in fact (\ref{zeta}) in which the first equation is (\ref{eqXi1}) with the thermodynamic potential (\ref{vdWn}) and the second equation is (\ref{vdWkex}) with the thermodynamic potential (\ref{vdWf}). In this formulation the time evolutions of $\nu(\rr)$ and $f(\rr,\vv)$ are not coupled but they are not the formulation  (\ref{clhi}) in which the time evolution governed by the second equation precedes the time evolution governed by the first equation.

In order to arrive at the  formulation  (\ref{clhi}), we make a second step in which we
combine the time evolution of $\nu(\rr)$ with the time evolution of $f(\rr,\vv)$  (i.e. write a new version of (\ref{zeta}) in which the time evolutions in the two equations are coupled) and then, by analyzing their solutions,  rewrite them into the form (\ref{clhi}). To begin this process we make the following one-to-one transformation $(\nu(\rr), f(\rr,\vv))\leftrightarrow (n(\rr),\phi(\rr,\vv))$
\begin{eqnarray}\label{trnnu}
n(\rr)&=&\vv(\rr)+\int d\vv f(\rr,\vv)\nonumber \\
\phi(\rr,\vv)&=&f(\rr,\vv)
\end{eqnarray}
The Poisson bracket (\ref{PBk}) is transformed by (\ref{trnnu}) into
another  Poisson bracket
\begin{eqnarray}\label{trr2}
\{A,B\}&=&\int d\rr\int d\vv \phi \left(\frac{\partial \frac{\partial A}{\partial \phi}}{\partial \rr} \frac{\partial \frac{\partial B}{\partial \phi}}{\partial \vv} - \frac{\partial \frac{\partial B}{\partial \phi}}{\partial \rr} \frac{\partial \frac{\partial A}{\partial \phi}}{\partial \vv}\right)\nonumber \\
&&+ \int d\rr\int d\vv \phi \left(\frac{\partial \frac{\partial A}{\partial n}}{\partial \rr} \frac{\partial \frac{\partial B}{\partial \phi}}{\partial \vv} - \frac{\partial \frac{\partial B}{\partial n}}{\partial \rr} \frac{\partial \frac{\partial A}{\partial \phi}}{\partial \vv}\right)
\end{eqnarray}
expressing kinematics of $(n(\rr),\phi(\rr,\vv))$.
In  calculations involved in the passage from (\ref{PBk}) to (\ref{trr2}) we use
$\frac{\partial}{\partial \nu(\rr)}\rightarrow \frac{\partial}{\partial n(\rr)}$, $\frac{\partial}{\partial f(\rr,\vv)}\rightarrow \frac{\partial}{\partial \phi(\rr,\vv)}+\frac{\partial}{\partial n(\rr)}$.
With the state variables $(n(\rr),\phi(\rr,\vv))$ and the Poisson bracket (\ref{trr2}) the GENERIC equation (\ref{generic}) becomes
\begin{eqnarray}\label{TRR}
\frac{\partial n}{\partial t}&=& -\frac{\partial}{\partial \rr}\left(\int d\vv\phi\frac{\partial \frac{\partial \widetilde{\Phi}}{\partial \phi}}{\partial \vv}\right)\nonumber \\
\frac{\partial \phi}{\partial t}&=&  -\frac{\partial}{\partial \rr}\left(\phi\frac{\partial \frac{\partial \widetilde{\Phi}}{\partial \phi}}{\partial \vv}\right) + \frac{\partial}{\partial \vv}\left(\phi\frac{\partial \frac{\partial \widetilde{\Phi}}{\partial \phi}}{\partial \rr}\right)
+ \frac{\partial}{\partial \vv}\left(\phi\frac{\partial \frac{\partial \widetilde{\Phi}}{\partial n}}{\partial \rr}\right)
\end{eqnarray}
This formulation equipped in addition with an appropriate dissipation is the hierarchy reformulation (\ref{zeta}) of (\ref{Geq})  which
is the point of departure of the analysis leading eventually to the formulation (\ref{clhi}). We note that (\ref{TRR}) as well as (\ref{Geq}) are both particular realizations of the GENERIC equations ({\ref{generic}).
The thermodynamic potential $\widetilde{\Phi}(n,\phi)$ as well as   the dissipation potential that enters in the added dissipative term have to emerge from the analysis of solutions to (\ref{Geq}) (or (\ref{Geq1})). The distribution function $\phi(\rr,\vv)$ has in (\ref{TRR}) the role of an extra state variable providing details that are not seen when only  $n(\rr)$ plays the role of the state variable. If the external forces are present then they enter the second equation in (\ref{TRR}) governing the time evolution of $\phi(\rr,\vv)$.

\subsection{Enskog Vlasov kinetic theory}\label{vdW3}

Investigation of the time evolution of the vdW gas has been initiated by de Sobrino \cite{deS}. The Enskog Vlasov  kinetic equation (EV equation) as an equation describing the time evolution of the vdW gas
appeared first in \cite{EVG}. The EV equation is   the point of departure  for  recent  investigations of dynamics of phase transitions \cite{Fraz}, \cite{Ben}, \cite{StrEV} \cite{Karlin}. In the rest of this section we  focus attention on  difficulties that  the EV kinetic equation has with proving its compatibility with the  vdW equilibrium thermodynamics. We  argue that the kinetic equation (\ref{vdWkeq}), that is manifestly compatible with the vdW equilibrium thermodynamics and possesses the GENERIC structure,  is its simpler and physically meaningful alternative.

The long range attractive potential is taken into account  in the EV equation in the same way as in (\ref{Geq}) but the hard core repulsion is treated in a different way. There are essentially four ways to include the hard core potential in dynamics. The first is in the energy. Due to the highly singular character of the potential this is feasible only in computer numerical simulations of particles. The second is in the entropy. This has been done in the previous section. The third way is in constraining the motion. The particles cannot be closer to each other than the hard core allows. Such constraints should  be expressed in a modification of the Poisson bivector $L$. The fourth way is replacing the Boltzmann collision term with the Enskog collision term. While the third way remains to be explored, the fourth way has been explored.
The particles in the Boltzmann  collisions are points, collisions take place at one point,  and the position vectors of  incoming and outgoing particles do not change. The particles in the Enskog collisions are hard spheres of diameter $b$ and not only momenta but also position vector of the particles involved in collisions change.

We introduce the following shorthand notation: $(\rr,\vv)=X; (\rr_1,\vv_1)=Y; (\rr',\vv')=X'; (\rr'_1,\vv'_1)=Y'; (X,Y)=Z;  (X',Y')=Z'$. The Enskog collision is a one-to-one transformation $(X,Y)\leftrightarrow (X',Y')$ in which the momentum and the kinetic energy are conserved (i.e. $\vv +\vv_1=\vv' +\vv'_1$ and $\vv^2+(\vv')^2=\vv_1^2+(\vv'_1)^2$). In the Boltzmann collisions $\rr=\rr_1=\rr'= \rr'_1$. The Enskog collisions enter the EV equation in the form
\begin{eqnarray}\label{Enskog}
\mathbb{E}\mathbb{V}(f)&=&\int dY\int dZ'\left(W(n,Z',Z)(f(X')f(Y')-W(n,Z,Z')f(X)f(Y)\right)\nonumber \\ &&=\mathbb{E}\mathbb{V}_s(f)+\mathbb{E}\mathbb{V}_a(f)\nonumber \\
\mathbb{E}\mathbb{V}_s(f)&=&\int dY\int dZ'\left(W_s(n,Z',Z)(f(X')f(Y')-W_s(n,Z,Z')f(X)f(Y)\right)\nonumber \\
\mathbb{E}\mathbb{V}_a(f)&=&\int dY\int dZ'\left(W_a(n,Z',Z)(f(X')f(Y')+W_a(n,Z,Z')f(X)f(Y)\right)\nonumber \\
\end{eqnarray}
where $W$ is symmetric with respect to $X\leftrightarrow Y$, $W=0$ unless the momentum and the kinetic energy are conserved, otherwise  $W>0$. The dependence $W$ on $n(\rr)$ is nonlocal. It suffices to  assume that the dependence is weakly nonlocal in the sense that $W$ depends on $n(\rr)$ and also on $\frac{\partial n(\rr)}{\partial \rr}$. The quantities $W_s$ and $W_a$ are defined by
\begin{eqnarray}\label{Tr2}
W_s(n,Z,Z')&=&\frac{1}{2}(W(n,Z,Z')+W(n,Z',Z))\nonumber \\
W_a(n,Z,Z')&=&\frac{1}{2}(W(n,Z,Z')-W(n,Z',Z))
\end{eqnarray}

Explicit formulation of the Enskog collision term (\ref{Enskog}) (that can be found in \cite{Fraz}, \cite{Ben}, \cite{StrEV} \cite{Karlin})
is not necessary for presenting the problem that the EV equation has with proving its compatibility with the vdW equilibrium thermodynamics. The essential difference between the Enskog collision term and the Boltzmann collision term is that the antisymmetric part $W_a$ is absent in the latter.

The EV kinetic equation
\begin{eqnarray}\label{EVeqq}
\frac{\partial f}{\partial t}&=&-\frac{\partial}{\partial \rr}\left(\frac{\vv}{m}f(\rr,\vv)\right)
+\mathbb{E}\mathbb{V}_{a}(f)
+\frac{\partial f(\rr,\vv)}{\partial\vv}\frac{\partial}{\partial\rr}\int d\rr_1 V_{pot}(|\rr-\rr_1|)n(\rr_1)\nonumber \\
&&+ \mathbb{E}\mathbb{V}_{s}(f)
\end{eqnarray}
 is Eq.(\ref{Geq}) in which the second term on its right hand side is replaced by $\mathbb{E}\mathbb{V}_a(f)$. Without the Vlasov term and $\mathbb{E}\mathbb{V}_a(f)$ the EV equation (\ref{EVeqq}) is the Boltzmann equation with a nonlocal Boltzmann collision term $\mathbb{E}\mathbb{V}_s(f)$ replacing the standard local Boltzmann collision term. This modified Boltzmann equation is a particular realization of the GENERIC equation (\ref{generic}) and is thus compatible with equilibrium thermodynamics but not with the vdW equilibrium thermodynamics with the vdW fundamental thermodynamic relation. The same is true for the Boltzmann Vlasov equation which is the EV equation (\ref{EVeqq}) without the term $\mathbb{E}\mathbb{V}_a(f)$.

 The term $\mathbb{E}\mathbb{V}_a(f)$ on the right hand side of (\ref{EVeqq}) contributes  to the non dissipative the time reversible part of the time evolution. The problem of casting the EV equation (\ref{EVeqq}) to the form of the GENERIC equation (\ref{generic}) remains open. It has been only shown in \cite{GrCan} that the vdW thermodynamic potential (\ref{vdWf}) plays the role of the Lyapunov function provided an extra assumption relating   $\mathbb{E}\mathbb{V}_a(f)$   to $k_Bf(\rr,\vv)\ln(1-bn(\rr)))$ is made.

  On the other hand, the kinetic equation (\ref{Geq}) in which  the hard core repulsion force is treated as an entropic force (as it is done in the vdW vwn der Waals van Kampen equilibrium theory) is a more straightforward and a simpler way to account for it in the time evolution. Because of the appearance of the temperature in the entropic force, the validity of treating the hard core force as an entropic force also in dynamics can be tested   (as it is done in the case of the entropic versus the metal elasticity) by comparing predictions based on (\ref{Geq}) with   the temperature dependence of experimentally observed dynamic behavior.

\section{Concluding Remarks}

Nonequilibrium thermodynamics investigates time evolution in the state space  that is harmonious with equilibrium thermodynamics. Existence of the approach to equilibrium is the zero law of thermodynamics. The first law is the conservation of energy, and the second law is the dynamical version of the maximization of entropy (the approach to equilibrium is driven by the gradient of entropy). In this paper we follow a parallel line  in the time evolution  that takes place in vector fields instead of   in the state space. On the parallel line the zero law is the existence of the approach to reduced mesoscopic levels, the second law is the dynamical version of the Onsager variational principle (the approach to mesoscopic levels is driven by the gradient of Rayleighian).  In the context of hierarchical reformulations of mesoscopic  time evolution equations (e.g.
Grad hierarchy) the rate dynamics and thermodynamics is the  lower part of the hierarchy. It provides the closure that makes the upper part of the hierarchy (representing the reduced dynamics) self contained and the fundamental relation of rate thermodynamics.

We are exploring relations between thermodynamics and rate thermodynamics. In particular, we show that the  rate entropy is in externally unforced systems closely related to the  rate of the entropy and that the critical behavior
manifesting itself in the rate thermodynamics as an increase of fluctuations manifests itself in the reduced time evolution as a bifurcation. Some of these results of a general nature are illustrated in the
investigation of dynamics of the van der Waals  fluid. In the illustration we extend the  van der Waals and van Kampen equilibrium analysis of the van der Waals  gas  to dynamics. In particular, following van der Waals and van Kampen, we treat in dynamics the hard core repulsive force as an entropic force. Solutions of the  kinetic equation introduced in this paper are guaranteed to agree with the van der Waals equilibrium thermodynamics.
\\
\\
\\

\textbf{Acknowledgement}
\\

I would like to thank O\v{g}ul Esen, V\'{a}clav Klika and  Michal Pavelka  for stimulating discussions.
\\

\end{document}